\documentclass[a4paper]{article}
\usepackage{ISCSLP2024}
\usepackage{ifthen}
\newboolean{blind}
\setboolean{blind}{false} 
\usepackage{stfloats}
\title{The NPU-HWC System for the ISCSLP 2024 Inspirational and Convincing Audio Generation Challenge }
\name{
	\ifthenelse{\boolean{blind}}{Anonymous to ISCSLP}
	{Dake Guo$^1$, Jixun Yao$^1$, Xinfa Zhu$^1$, Kangxiang Xia$^1$, Zhao Guo$^1$, Ziyu Zhang$^1$, Yao Wang$^2$, Jie Liu$^2$, Lei Xie$^1$}
}
\address{
  \ifthenelse{\boolean{blind}}{Anonymous to ISCSLP}
  {
  	$^1$Audio, Speech and Language Processing Group (ASLP@NPU), School of Computer Science, \\ Northwestern Polytechnical University, Xi'an, China\\
  $^2$Huawei Cloud
  }
}

\email{
	\ifthenelse{\boolean{blind}}{Anonymous to ISCSLP}
	{guodake@mail.nwpu.edu, coauthor@company.com}
}

\begin{document}

\maketitle
\begin{abstract}
This paper presents the NPU-HWC system submitted to the ISCSLP 2024 Inspirational and Convincing Audio Generation Challenge 2024 (ICAGC). Our system consists of two modules: a speech generator for Track 1 and a background audio generator for Track 2.
In Track 1, we employ Single-Codec to tokenize the speech into discrete tokens and use a language-model-based approach to achieve zero-shot speaking style cloning. The Single-Codec effectively decouples timbre and speaking style at the token level, reducing the acoustic modeling burden on the autoregressive language model. Additionally, we use DSPGAN to upsample 16 kHz mel-spectrograms to high-fidelity 48 kHz waveforms.
In Track 2, we propose a background audio generator based on large language models (LLMs). This system produces scene-appropriate accompaniment descriptions, synthesizes background audio with Tango 2, and integrates it with the speech generated by our Track 1 system.
Our submission achieves the second place and the first place in Track 1 and Track 2 respectively.
\end{abstract}
\section{Introduction}

Deep learning technologies have significantly advanced speech synthesis, enabling the generation of high-quality and human-like speech~\cite{DBLP:conf/interspeech/WangSSWWJYXCBLA17,DBLP:conf/nips/RenRTQZZL19,DBLP:conf/icml/KimKS21}.
Recent breakthroughs, such as language models and diffusion models~\cite{shen2023naturalspeech,DBLP:journals/corr/abs-2305-07243}, powered by massive datasets and extensive model parameters, have pushed the capabilities of speech synthesis systems to new heights, making it possible to produce speech that matches human-level performance.

To generate highly realistic and synthetic speech close to natural sound, it not only needs to generate high-quality vocals but also needs to incorporate appropriate background sounds. These background sounds serve to add layers of realism and context, significantly enhancing the overall auditory experience.
Expressive speech ensures that the synthetic audio is engaging, conveying the intended styles and nuances of the speaker effectively. Additionally, background sounds also play a crucial role in improving the realism and context of the synthetic speech. They provide an auditory environment that can place the speech in a specific setting, making it more realistic.

To promote the development of realistic speech synthesis, the International Conference on Chinese Spoken Language Processing 2024 (ISCSLP 2024) has initiated the Inspirational and Convincing Audio Generation Challenge 2024 (ICAGC 2024)~\cite{icagc2024}. This challenge focuses on generating high-quality audio close to real-world sound. The challenge is divided into two tracks. Track 1 involves a zero-shot speaker and style cloning, where participants use a speech sample from a target speaker to clone the voice and modulate it to convey various themes, such as novel chapters and ancient Chinese poems. Track 2 focuses on background audio generation. Participants are tasked with adding background sounds to the synthetic speech from Track 1, accurately representing real-life scenarios,  thereby enhancing the realism of the audio.
\begin{figure*}[hbt]
  \centering
  \includegraphics[width=0.85\linewidth]{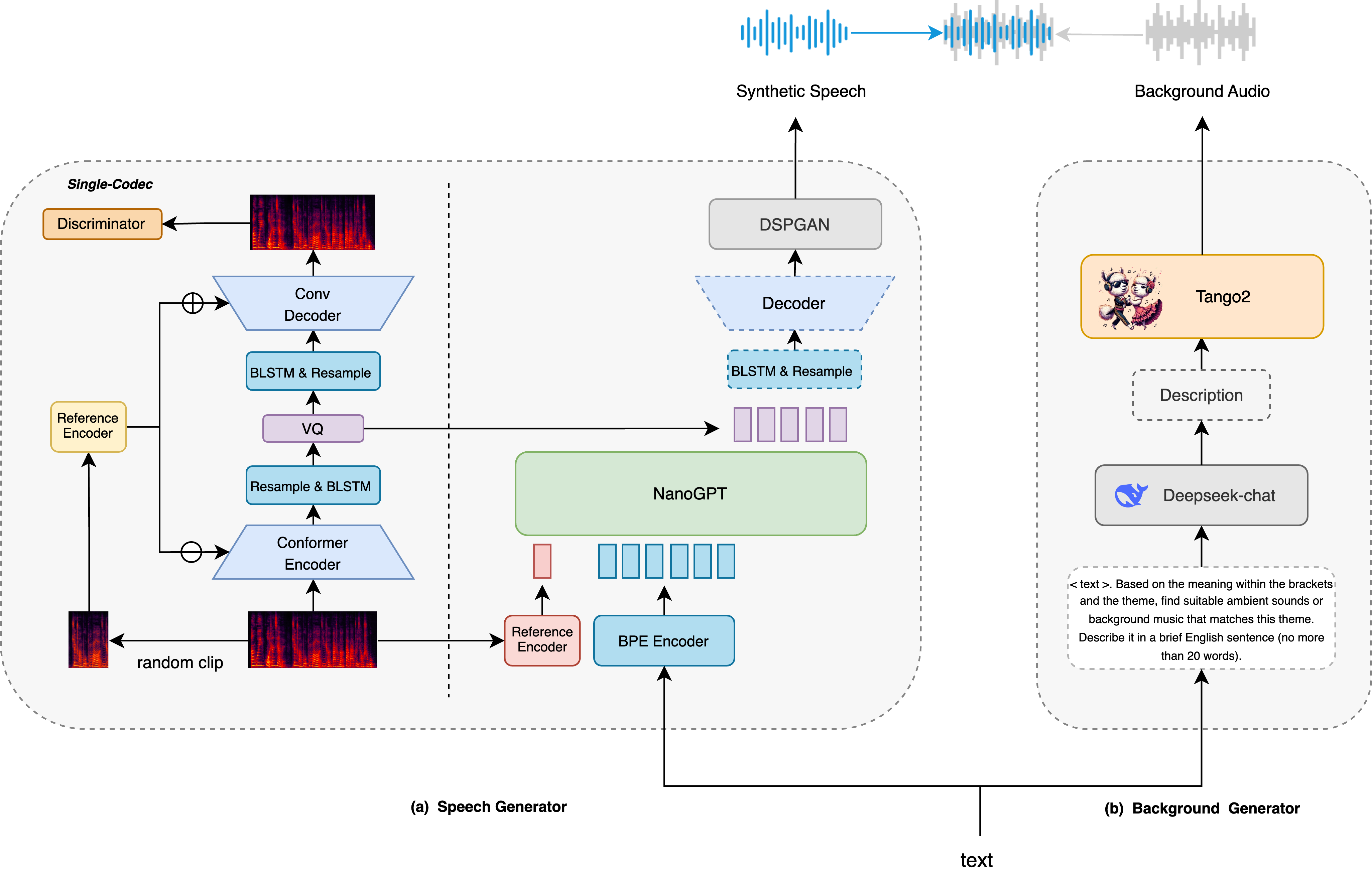}
  \caption{Overall architecture of our submitted background audio and speech generation system}
  \label{fig:overall}
  \vspace{-10pt}
\end{figure*}


This paper describes our systems developed for both tracks in the ICAGC 2024 challenge. In Track 1, we implement a zero-shot speaker and style cloning speech generator. Specifically, we use a Single-Codec~\cite{li2024single} as our speech tokenizer, which explicitly decouples timbre and speaking style at the token level, thereby reducing the acoustic modeling burden on the autoregressive model. Moreover, we adopt the DSPGAN~\cite{10095105} vocoder to upsample 16 kHz mel-spectrograms to high-fidelity 48 kHz waveforms.
In Track 2, we introduce a cascaded background audio generation system based on large language models (LLMs), which generate scene-appropriate accompaniment descriptions. Additionally, it synthesizes background audio with Tango 2 and mixes it with the vocal speech produced by the Track 1 system. Our proposed system achieves an average Mean Opinion Score (MOS) of 3.63 in Track 1, ranking the second place, and an average MOS of 3.67 in Track 2, securing the first place.

\section{Related Work}

\subsection{Zero-shot Speaker and Style Cloning}

Zero-shot and style speaker cloning aims to generate speech in the same style as an unseen speaker, given only a few seconds of acoustic reference. Speaker modeling-based approaches have been commonly used to achieve zero-shot speaker cloning. Several works have focused on extracting a global speaker embedding from a given reference speech using a pre-trained speaker verification~(SV) model or a joint-trained speaker encoder~\cite{yourtts,casanova2021sc}. However, the limited information-capturing capability of a single vector in these global speaker modeling approaches often results in less resemblance between synthetic speech and the speaker’s original voice. To address this, other works have aimed to capture both global timbre information and local prosodic variations through multi-level speaker representation, thereby improving speaker similarity for zero-shot speaker and style cloning~\cite{fu2019phoneme,zhou22d_interspeech,li2022unet, vectok}.


Recent breakthroughs in speech synthesis have significantly enhanced speech realism by leveraging massive datasets and extensive model parameters. NaturalSpeech 2~\cite{shen2023naturalspeech} introduced a speech prompting mechanism that encourages predicted duration and pitch to follow diverse speaker information from the speech prompt. Mega-TTS 2~\cite{jiang2023mega} uses a multi-reference timbre encoder to extract timbre information from multiple reference speeches.
Additionally, some works employ in-context learning powered by language models to achieve zero-shot speaker and style cloning. For instance, VALL-E~\cite{DBLP:journals/corr/abs-2301-02111} achieves voice cloning with just a 3-second prompt through in-context learning. Tortoise TTS~\cite{DBLP:journals/corr/abs-2305-07243} and XTTS~\cite{DBLP:journals/corr/abs-2406-04904}, utilizing the Perceiver model~\cite{DBLP:conf/nips/AlayracDLMBHLMM22}, accomplish voice cloning with a single reference audio from the target speaker. BASE TTS~\cite{DBLP:journals/corr/abs-2402-08093} extends the capabilities to larger datasets and model scales, enhancing the proficiency in presenting the appropriate prosody for complex texts. Seed-TTS~\cite{DBLP:journals/corr/abs-2406-02430} further demonstrates the potential of large model architectures by synthesizing speech that matches human levels in naturalness and expressiveness. These advancements harness the robust semantic processing power of large models to produce highly expressive speech based on textual inputs and acoustic prompts.
\subsection{Text to Audio}
Text to Audio (TTA) aims to generate audio, including environmental sounds and background music, based on textual descriptions. TTA can potentially be used to enhance synthetic speech by providing a more immersive and contextually appropriate auditory experience. AudioGen~\cite{DBLP:conf/iclr/KreukSPSDCPTA23} treats TTA as a conditional language modeling task, leveraging the contextual nuances captured by language models to generate cohesive audio scenes. On the other hand, models like AudioLDM~\cite{DBLP:conf/icml/LiuCYMLM0P23} and Make-An-Audio~\cite{DBLP:conf/icml/HuangHY0LLYLYZ23} adopt latent diffusion techniques, showcasing their unique approaches to audio generation. Tango 2~\cite{DBLP:journals/corr/abs-2404-09956} has further advanced TTA performance by implementing Direct Preference Optimization (DPO)~\cite{DBLP:conf/nips/RafailovSMMEF23} within its latent diffusion framework.

\section{Methodology}
As illustrated in Figure~\ref{fig:overall}, our system is structured into two main components: the speech generator and the background audio generator. The speech generator is engineered to produce speech that matches the timbre and style of the target speaker. Meanwhile, the background audio generator is designed to create appropriate background sounds based on the speech transcripts. This dual-component system ensures that the synthetic speech is not only realistic in terms of timbre similarity but also enriched with contextual sounds.

\subsection{Zero-shot Speaker and Style Cloning}


As shown in Figure~\ref{fig:overall} (a), the speech generator comprises four main components: text and speech tokenizers for encoding different modal information, a language model for generating audio tokens, and a vocoder for reconstructing audio from mel-spectrograms.

\subsubsection{Tokenizer}
We adopt Byte Pair Encoding (BPE) as our text tokenizer rather than traditional phonemes. BPE works by iteratively merging frequently occurring characters or sequences of characters to form new encoding units, resulting in a more compact and efficient text representation. This approach helps not only to minimize cumulative errors in text front-end processing but also to capture the semantic information in the text better.

For speech tokenization, we utilize the Single-Codec. Single-Codec targets the mel-spectrogram for modeling, effectively compressing speech information while preserving essential details. It employs a decoupled Vector Quantized Variational Autoencoder (VQ-VAE), segmenting the speech into timbre, acoustic environment-related time-invariant embeddings, and a discrete sequence related to pronunciation. Additionally, by integrating a BLSTM module, mixed sampling, and resampling techniques, the Single-Codec enhances the efficiency of speech content modeling and facilitates the formation of stable clustering units. With fewer but richer phonetic discrete representations, Single-Codec alleviates the modeling burden on the language model, enabling it to reconstruct highly expressive audio at lower bit rates.

\subsubsection{Language Model}
Given text-speech pairs $\{\mathbf{x}_i,\mathbf{s}i\}^{N}_{i=0}$, where $\mathbf{x}$ represents the text transcription and $\mathbf{s}$ denotes the corresponding audio sample, we process these pairs to generate speech tokens. The text token $\overline{\mathbf{x}}$ is obtained by the BPE tokenizer, while the speech token $\overline{\mathbf{s}}$ is extracted by the Single-Codec from the speech $\mathbf{s}$.
Similar to previous work, we employ a Language Model (LM)-based approach for Text-to-Speech (TTS) generation.
We use a NanoGPT-architecture\footnote{https://github.com/karpathy/nanoGPT} as a language model, parameterized by $\theta$, to predict speech tokens conditioned on both text and reference speech $\mathbf{s}_{ref}$ in an autoregressive manner.  The reference speech $\mathbf{s}_{ref}$ involves a randomly selected mel-spectrum clip from the target speech $\mathbf{s}$ and then encoded into a fixed-size embedding $\mathbf{e}_{\text{ref}}$. The reference embedding $\mathbf{e}_{\text{ref}}$, alongside the text token $\overline{\mathbf{x}}$ and speech token $\overline{\mathbf{s}}$, are concatenated into a single sequence and employ a next-token prediction modeling by the GPT-based autoregressive model. To enhance the capability to distinguish and accurately process the various types of input, we employ distinct positional embeddings for text tokens $\mathbf{t}$ and speech tokens $\mathbf{s}$.

At the training stage, we employ the teacher-forcing scheme, in which the left-shifted sequence is employed as the mode inputs and the original sequence serves as the expected outputs. In order to retain textual information to guide prosody, the cross-entropy loss is calculated on both text token $\overline{\mathbf{x}}$ and speech token $\overline{\mathbf{s}}$:
{
\begin{align*}
\mathcal{L} = &-\sum_{n=0}^{N} \left(\log P\left(\overline{\mathbf{x}}_{n} \mid \overline{\mathbf{x}}_{<n}, \mathbf{e}_{\text{ref}} ; \theta\right) \right) \\
&- \sum_{m=0}^{M} \left(\log P\left(\overline{\mathbf{s}}_{m} \mid \overline{\mathbf{x}}_{<N}, \overline{\mathbf{s}}_{<m}, \mathbf{e}_{\text{ref}} ; \theta\right)\right) 
\end{align*}
}
Where $N$ is the length of the text sequence and $M$ is the length of the speech sequence.



\subsubsection{Vocoder}
We implement frequency band expansion in the vocoder to enhance speech quality, particularly under conditions where the primary training data comprises low-resolution 16 kHz audio. We utilize DSPGAN~\cite{10095105} to upsample the 16 kHz mel-spectrogram to a high-fidelity 48 kHz waveform which effectively bridges the gap between the inherent limitations of the training data and the demand for high-quality speech generation.

\subsection{Background Audio Generation}
As shown in Figure~\ref{fig:overall} (b), we develop a LLM-based system to generate appropriate background audio for synthetic speech, consisting of two main stages: first analyze the speech text to generate background sound description (Text-to-Description) and then synthesize background audio based on the generated description (Description-to-Audio).

\subsubsection{Text-to-Description}
In the Text-to-Description stage, the goal is to generate descriptive texts of potential scenes corresponding to the speech transcripts using the semantic understanding capabilities of LLM. For this purpose, we choose DeepSeek-V2\footnote{https://github.com/deepseek-ai/DeepSeek-V2}, a robust Mixture-of-Experts (MoE) model known for its strength in language understanding. However, we are facing two significant challenges. First, the LLM tends to produce complex and lengthy scene descriptions, which is problematic as  TTA technology performs better with simpler descriptions. Second, the LLM could generate appropriate descriptions when the text involves concrete objects, such as ancient poems. However, in cases where the text lacks concrete elements or has very few (such as in philosophical texts like the Analects), the LLM's generated scene descriptions tend to deviate from relevance, as texts without tangible objects inherently lack a clear corresponding scene.

To address these challenges, we refine our prompting strategy for the LLM, as shown in the right of figure \ref{fig:overall}. We limit the complexity of the descriptions in the prompts and provide the LLM with two options: to generate either a scene or a musical description. This ensures that the background audio generated can effectively match and enhance both concrete and abstract texts, making the synthetic output more versatile and contextually appropriate.

\subsubsection{Description-to-Audio}

We choose \textit{Tango 2}~\footnote{https://github.com/declare-lab/tango} to generate background sounds. Tango 2 is an advanced TTA model that combines a latent diffusion model, a textual-prompt encoder powered by the FLAN-T5~\cite{DBLP:journals/corr/abs-2210-11416} large language model. It introduces unique features such as the Audio-alpaca dataset, designed for preference alignment, and employs diffusion-DPO for training, which helps in producing audio outputs that closely match human perceptual preferences. These methods make Tango 2 particularly effective in creating highly realistic and contextually aligned audio from text prompts. We convert the scene or musical descriptions obtained from the Text-to-Description stage into corresponding audio. Furthermore, because the primary focus of the audio is vocal content, to ensure that the background audio does not overpower the main speech, we apply a 10dB gain reduction to the generated background audio during mixing. 
\section{Experiments}

\subsection{Dataset}

We use over 80,000 hours of Mandarin speech data in our training set. This includes 10,000 hours of open-source data from resources like WenetSpeech4TTS~\cite{wenetspeech4tts} and DiDiSpeech~\cite{DBLP:conf/icassp/GuoWJLZZLGZHL21}. The remaining 70,000 hours of data are sourced from Chinese podcasts and audiobooks available online. To obtain high-quality audio segments from single speakers, we employ the same data processing pipeline as it in WenetSpeech4TTS~\cite{wenetspeech4tts}.

\subsection{Training Settings}
We use \textit{r50k\_base} with tiktoken\footnote{https://github.com/openai/tiktoken} from OpenAI for text processing, which features a vocabulary size of 50,257. Our Single-Codec, NanoGPT, and DSPGAN models are each trained independently on distinct datasets tailored to their specific requirements. The Single-Codec model is trained following the protocols outlined in its original paper, but we scale up the dataset, incorporating over 80k hours of audio clips. DSPGAN is also trained according to its original setup, but it leverages an internal high-fidelity dataset of approximately 1k hours to ensure it produces high-quality audio outputs. NanoGPT is configured with about 400M parameters, encompassing 24 transformer layers, each with 16 heads. The input embedding dimension for NanoGPT is set to 1024.

The training setup for our models involves substantial computational resources and specific configurations to optimize performance. The Single-Codec model is trained on 8 RTX 4090 GPUs, using a batch size of 1024 across 300k steps. In contrast, DSPGAN is trained on 4 RTX 4090 GPUs with a batch size of 256, undergoing 300k iterations to achieve high fidelity in audio generation. NanoGPT, which requires a more extensive training regimen due to its complexity, is trained on 8 RTX A6000 GPUs. It uses a much smaller batch size of 6 over 1,400k steps to ensure detailed attention to the speech synthesis process. Additionally, to manage the training effectively with such a small batch size, we employ a gradient accumulation strategy, accumulating gradients 100 times before updating the model parameters. 
\subsection{Evaluation metrics}
We evaluate the performance of Track 1 on the reserved test set with speaker embedding cosine similarity (SECS) and character error rate (CER) in ASR.  The test set comprised audio samples from 20 different speakers, with each speaker generating 430 identical sentences to maintain consistency across the evaluations. The SECS metric is computed by extracting speaker embeddings with Resemblyzer~\footnote{https://github.com/resemble-ai/Resemblyzer} and calculating the cosine similarity. The CER is measured between transcripts of real and synthesized utterances transcribed by Paraformer~\cite{gao2022paraformer}.

In the official competition, the evaluation metrics include Quality Mean Opinion Score (MOS), Similarity MOS, Emotion MOS, and Background Audio Matching Degree MOS. The Quality MOS measures the overall perceptual quality of the synthesized speech. The Similarity MOS assesses how closely the synthesized speech mirrors the vocal characteristics of the target speaker. The Emotion MOS evaluates the synthetic speech's effectiveness in conveying intended emotions that resonate and inspire listeners on a cognitive level, where native speakers judge the emotional impact and inspirational quality of the speech. Finally, the Audio and Speech Convincing Matching Degree metric gauges how well audio enhancements like sound effects and background sounds integrate with the speech, enhancing the overall realism. 
\subsection{Track 1}
We conduct a comparative analysis between our speech generator and an open-source system, Chinese VALL-E~\footnote{https://github.com/dukGuo/valle-audiodec} trained on WenetSpeech4TTS.
As shown in Table 1, our system demonstrates outstanding performance in both CER and SECS. It achieved a CER of 6.86\%, significantly lower than VALL-E, indicating a substantial advantage in reducing character errors and enhancing speech accuracy. Additionally,  our system achieves an SECS of 0.849, showcasing strong competitiveness, far surpassing VALL-E. 
As shown in Table 1, our system outperforms the open-source system in terms of CER. This superiority in speaker similarity demonstrates that our system not only achieves a closer match to the target speaker's voice but also enhances the stability of the auto-regressive model, resulting in fewer transcription errors. 
\begin{table}[hbt]
\centering
\caption{Objective evaluations results for speech generator}
\resizebox{0.6\linewidth}{!}{
\begin{tabular}{ccc}
\toprule
\textbf{Method} & \textbf{CER} $\downarrow$   & \textbf{SECS} $\uparrow$ \\ \midrule
NPU-HWC (ours)  & \textbf{6.86}\% & 0.849         \\ 
VALL-E  &13.10\% & 0.812        \\ \bottomrule
\end{tabular}
}
\vspace{-6pt}
\end{table}

In the official listening tests, our system ranks second in average Mean Opinion Score (MOS) as shown in Table 2. Our use of DSPGAN to transform 16 kHz mel-spectrograms into high-fidelity 48 kHz waveforms allows us to achieve a Quality MOS that is closely competitive with the top-ranked system. However, our speaker similarity score is relatively lower due to our reliance on a single frame of speaker embedding, which may limit the ability to capture the full range of the speaker's timbre nuances. Despite not achieving the highest score in Emotion MOS, our system still outperformed most other teams and ranks just below the first place. This success can be attributed to our strategy of decoupling timbre at the token level, which significantly enhanced the language model’s capacity for semantic modeling.
\vspace{-8pt}
\begin{table}[hbt]
\centering
\caption{ Evaluation results for Track 1}
\resizebox{0.9\linewidth}{!}{
\begin{tabular}{ccccc}
\toprule
\textbf{Team} & \textbf{Quality} $\uparrow$   & \textbf{Similarity} $\uparrow$  & \textbf{Emotion} $\uparrow$  & \textbf{Average} $\uparrow$\\ \midrule
YouCaiHua  & \textbf{3.89} & \textbf{3.83}   & \textbf{3.85}  &\textbf{ 3.86 }  \\ 
NPU-HWC (ours)  &3.84 & 3.48   & 3.58     & 3.64  \\ 
zyzx\_ai  &3.67 & 3.50  & 3.50     & 3.56  \\ 
PeiyangTTSer  &3.64 & 3.52  & 3.28     & 3.48  \\ 
Happy Happy &3.33 & 3.45   & 3.33    & 3.37  \\ \bottomrule
\end{tabular}
}
\vspace{-10pt}
\end{table}

\subsection{Track 2}
As shown in Table 3, in the subjective tests for Track 2, our system outperforms another team in all aspects. A significant factor contributing to our success is the use of a LLM to generate scene or music descriptions that better match the transcripts.
\vspace{-8pt}
\begin{table}[hbt]
\centering
\caption{ Evaluations result for Track 2}

\resizebox{0.9\linewidth}{!}{
\begin{tabular}{ccccc}
\toprule
\textbf{Team}   & \textbf{Similarity} $\uparrow$  & \textbf{Emotion} $\uparrow$ & \textbf{Matching} $\uparrow$   & \textbf{Average} $\uparrow$\\ \midrule

NPU-HWC (ours)  &\textbf{4.15} &\textbf{ 3.66}   & \textbf{3.33}     & \textbf{3.67}  \\ 
PeiyangTTSer  &3.33 & 3.33  &2.66    & 3.06  \\ \bottomrule
\end{tabular}
}
\vspace{-10pt}
\end{table}

\section{Conclusion}
In this paper, we introduce background audio and speech generation system for IGAGC2 2024.
For Track 1, we created a speech generator capable of zero-shot speaking style cloning. We employed Single-Codec as the speech tokenizer, which decouples timbre and speaking style at the token level, thus reducing the acoustic modeling load on the autoregressive model. Additionally, we used the DSPGAN vocoder to upsample 16 kHz mel-spectrograms to high-fidelity 48 kHz waveforms.
In Track 2, we employ a cascading system to generate background audio using LLM. This process begins with the LLM creating scene-appropriate accompaniment descriptions. Next, these descriptions are used by Tango 2 to synthesize the background audios. Finally, the synthesized background audio is mixed with the vocal audio produced in Track 1.
Our system achieves an average Mean Opinion Score (MOS) of 3.63 in Track 1, and an even higher average MOS of 3.67 in Track 2.

\bibliographystyle{IEEEtran}

\bibliography{mybib}


\end{document}